\documentclass[aps,prd,preprint,groupedaddress]{revtex4-1}

\begin{document}


\title{Gauge mediation with heavy doublet superparticles}


\author{Hua.~Shao}
\email[]{shaohua@itp.ac.cn}
\affiliation{State Key Laboratory of Theoretical Physics,
Institute of Theoretical Physics, Chinese Academy of Sciences,
P.O. Box 2735, Beijing 100190, P.R. China}


\date{\today}

\begin{abstract}
It is challenging for supersymmetry if the 125~\textmd{GeV} Higgs boson is confirmed by the LHC. In the case of small squark mixing it is inevitable to introduce heavy top squarks to lift the Higgs mass that is hard to be produced by the LHC. Here we consider the possibility that in gauge mediation the superparticles belonging  to $SU(2)$  doublets are much heavier than those do not carry the $SU(2)$ quantum numbers. Under the assumption not only the Higgs mass can be large enough but also there are light right handed top squarks below the \textmd{TeV} scale that can be observed in future.
\end{abstract}


\maketitle

\section{Introduction}
Supersymmetry is one of the most favored candidates for new physics beyond the Standard Model~\cite{wess and bagger,Nilles:1983ge}. One of the main motivation for building the Large Hadron Collider~(LHC) at CERN is to find the supersymmetric particles. If supersymmetry really describes nature the recent LHC results imply that the superpartners for known Standard Model particles, the sparticles, are heavy. It agrees with the Higgs mass if we treat the excess of events around $125~\textmd{GeV}$ in ATLAS~\cite{AtlasTalk} and CMS~\cite{CMSTalk} as the signals of the Higgs bosons. In the Minimal Supersymmetric Standard Model~(MSSM) heavy sparticles are needed to generate a heavy Higgs boson~\cite{Haber:1990aw,Okada:1990vk,Ellis:1990nz}. It is  challenging to theories, comes from the following aspects.

\emph{Observability}. A detailed calculation shows that in the absence of squark mixing the higgs boson can not be $125~\textmd{GeV}$ in a large area in the parameter space for the top squark lighter than $3~\textmd{TeV}$\cite{Hall:2011aa,Arbey:2011ab}. It will beyond the  the ability of the LHC to create them. It is significant to find a theory in which there are light colored sparticle that can be produced by the LHC and the there are still enough quantum corrections to the Higgs mass.

\emph{Naturalness}. Heavy top squrks will introduce large corrections to the Higgs mass parameter which proportional to their mass square
\begin{equation}
\delta m_h^{(t)2}\sim -\frac{3y_t^2}{4\pi^2}m_{\widetilde{t}}^2\log\frac{M}{m_{\widetilde{t}}},
\end{equation}
where $M$ is the new physics scale. There must be other contributions to the Higgs potential to compensate such a large corrections in order to break the electroweak symmetry at right scale. Under the present condition it is inevitable to do a fine-tuning to obtain the weak scale. In the usual theories with low energy supersymmetry breaking the fine-tuning is sensitive to the top squark mass. A natural question is that if there is a fine-tuned supersymmetric theory but the sensitivity of the fine-tuning to the top squark mass is lower.

\emph{Predictivity}. There are over 100 free parameters in the MSSM, most of them are soft supersymmetry breaking parameters. In principle they can be predicted if the underlying theory of supersymmetry breaking and mediation is known. It is important to find and identify the supersymmetry breaking and mediation mechanism.

Theories with gauge mediated supersymmetry breaking~\cite{Giudice:1998bp,Dine:1993yw,Dine:1994vc,Dine:1995ag} is attractive to predict these breaking parameters. There are only ten free parameters for the MSSM. Six of them are enough to determine all the soft breaking masses~\cite{Meade:2008wd} and the other four are parameters in the Higgs sector, soft masses of $H_{u,d}$, Dirac mass $\mu$ and its corresponding bilinear breaking $B\mu$~\cite{Komargodski:2008ax}. The $B\mu$ is generally much larger than the weak scale if there is no specified  structure in the Higgs sector~\cite{Dvali:1996cu}.

Recall that the Higgs mass matrix in gauge mediation is
\begin{equation}\label{higgs mass matrix}
M_H^2=\left(
        \begin{array}{cc}
         |\mu|^2+ m_{\widetilde{e}_L}^2+ \delta m_u^2+\delta m_h^{(t)2} & -B\mu \\
          -B\mu & |\mu|^2+m_{\widetilde{e}_L}^2+\delta m_d^2 \\
        \end{array}
      \right),
\end{equation}
where $m_{\widetilde{e}_L}^2$ is the gauge mediated left handed slepton mass.
Superparticles will obtain the same soft masses in gauge mediation if they have the same Standard Model quantum numbers.
The quantum numbers of the Higgs doublets $H_{u,d}$ are the same as the left handed sleptons up to the sign of the weak hypercharges, so the gauge mediated soft breaking masses for $H_{u,d}$ are $m_{\widetilde{e}_L}^2$.
$\delta m_{u,d}^2$ are the soft masses from  the mechanism which  generates $\mu$ and $B\mu$.
In eq.~(\ref{higgs mass matrix}) the  general estimation in theories with low energy supersymmetry breaking and mediation
\begin{equation}
\frac{B\mu}{\mu^2}\sim 16\pi^2\gg 1
\end{equation}
is assumed.
It is seen that both large $B\mu$ and heavy top squarks tend to make one of the eigenvalue of eq.~(\ref{higgs mass matrix}) to be negative with large absolute value. There must be large positive contributions to obtain the weak scale.

There are many excellent works on electroweak symmetry breaking under large $B\mu$~\cite{Csaki:2008sr,DeSimone:2011va}, here a new possibility is given. Assuming that there is a large hierarchy between sparticles according to their quantum numbers, superpartners of left-handed fields are much heavier than those are $SU(2)_L$ singlets. The large left-handed superparticles' mass are used to compensate the above two negative contributions. We will see that the electroweak symmetry breaking can happen under fine-tuning which is less dependent on the top squark mass, the right-handed squarks and sleptons remain light below the $\textmd{TeV}$ scale that will be observed in the future LHC experiment.

 The basic structure and predictions  of this scenario is given in the next section. In the last section, we will give  brief discussions and conclusions of this article.

\section{Doublet dominant gauge mediation}
For simplicity it is assumed that the parameters in the Higgs sector are generated by simple dynamics. For theories with low energy supersymmetry breaking $B\mu/\mu^2\gg1$. There must be fine-tuning for correct electroweak symmetry breaking.

The most general sfermion masses in gauge mediation are given by
\begin{equation}
m_{\widetilde{f}}^2=\sum_{r=1}^{3}(\frac{\alpha_i}{4\pi})^2C_r^{\widetilde{f}}A_r,
\end{equation}
where $C_r^{\widetilde{f}}$ is  the quadratic Casimir of the sparticles and $A_r$'s are the gauge mediated supersymmetry breaking effects which are determined by the correlation functions of the underlying dynamics~\cite{Meade:2008wd}. They have mass dimension 2.In principle there is no prior reason that the three $A_r$'s are of the same order of magnitude, they can be separated indeed.

Assuming that $A_2\gg A_{1,3}$, the sparticles of left-handed fermions are much heavier then others. The sfermion spectrum has the following patten
\begin{equation}\label{mass patten}
m_{\widetilde{Q}_L}^2\sim m_{\widetilde{e}_L}^2\gg m_{\widetilde{t}_R}^2\sim m_{\widetilde{b}_R}^2\sim m_{\widetilde{e}_R}^2.\end{equation}

The simplest mechanism  to realize such a patten is
\begin{equation}\label{gauge mediation details}
W=\sum_{r=1}^3 X_i\phi_i\overline{\phi}_i+\lambda_uH_u\phi_1\phi_2+\lambda_dH_d\overline{\phi}_1\overline{\phi}_2,
\end{equation}
where $\phi_i$ and $\overline{\phi}_i$ are messengers, fields with overlines are in the complex conjugation representation of those without overlines. The Standard Model $SU(2)_L\times U(1)_Y$ quantum numbers are $(0,-1)$ for $\phi_1$ and $(2,0)$ for $\phi_2$, respectively. The vacuum expectation values of the supersymmetry breaking fields $X_i$ are
\begin{equation}
\langle X_i\rangle=M_i+\theta^2F_i.
\end{equation}
For simplicity we assume that $M_1\sim M_2\sim M_3$ and $F_2\gg F_{1,3}$. It is easy to see that the superparticles in doublets are heavier than singlets. Under this patten the top squark correction to the Higgs mass matrix~(\ref{higgs mass matrix})
\begin{equation}\label{small correction of stop}
\delta m_h^{(t)2}/m_{\widetilde{e}_L}^2\sim  -\frac{3y_t^2}{8\pi^2}\log\frac{M}{m_{\widetilde{t}}}\ll 1,
\end{equation}
it is less important than other quantities.

The four parameters in the Higgs sector in the assumed limit are~\cite{DeSimone:2011va}
\begin{equation}
\delta m_{u,d}^2\simeq\frac{\lambda_{u,d}^2}{16\pi^2}(\frac{F_2}{M_2})^2,
\end{equation}
\begin{equation}\label{mu}
\mu\simeq\frac{\lambda_u\lambda_d}{16\pi^2}\frac{F_2}{M_2}
\end{equation}
and
\begin{equation}
B\mu\simeq\frac{\lambda_u\lambda_d}{16\pi^2}(\frac{F_2}{M_2})^2.
\end{equation}
The gauge mediated left-handed selectron $\widetilde{e}_L$ mass is
\begin{equation}
m_{\widetilde{e}_L}^2=2\cdot\frac{3}{4}(\frac{\alpha_2}{4\pi})^2(\frac{F_2}{M_2})^2+2\cdot(\frac{\alpha_1}{4\pi})^2(\frac{F_1}{M_1})^2\approx2\cdot\frac{3}{4}(\frac{\alpha_2}{4\pi})^2(\frac{F_2}{M_2})^2\approx m_{\widetilde{t}_L}^2.\end{equation}

To the first order that neglect contributions from the other two gauge interactions, the Higgs mass matrix~(\ref{higgs mass matrix}) proportional to $(\frac{F_2}{M_2})^2$. We can always choose $\lambda_{u,d}$ to make one of the eigenvalue of eq.~(\ref{higgs mass matrix}) is at the weak scale no matter how large $(\frac{F_2}{M_2})^2$ is. It is easy to see this fact. Two of the  eigenvalues of eq.~(\ref{higgs mass matrix}) are all positive for $\lambda_u=\lambda_d=0$ since $m_{\widetilde{e}_L}^2$ is always positive and the small top squark correction~(\ref{small correction of stop}). On the other hand for large $\lambda_{u,d}$ in which $\delta m_{u,d}^2\gg m_{\widetilde{e}_L}^2$ one of the eigenvalues is always negative with large absolute value which is of the order of $\frac{\lambda^2}{16\pi^2}(\frac{F_2}{M_2})^2$~\cite{DeSimone:2011va}. Due to the continuousness of the eigenvalues, there must be some $\lambda_{u,d}$ to make the negative eigenvalue is at the weak scale.

The most possible orders of magnitude of the couplings $\lambda_{u,d}$ can be inferred by the cancelation between the dynamical generated soft masses for the Higgs fields and the gauge mediated mediated heavy doublet masses. The requirement
\begin{equation}
 \delta m_{u,d}^2\sim m_{\widetilde{e}_L}^2
 \end{equation}
implies the estimation of order
\begin{equation}\label{estimation}
\lambda_{u,d}^2\sim\alpha_2^2.
\end{equation}
It can be seen that  the one loop generating soft masses for $H_{u,d}$ should be of the same order of the two loop generated soft masses for doublet sfermions. Since $\lambda_{u,d}$ are very small numbers,  large enough $F_2/M_2$ is generally required for phenomenological reasons.

Assume that in the absence of the contributions of the top squarks the above fine tuning has been performed in which for given $F_2/M_2$ the couplings $\lambda_{u,d}$ have been selected to make the correct electroweak symmetry breaking, for finite top squark mass $m_{\widetilde{t}_L}$ the element of the mass matrix~(\ref{higgs mass matrix}) have to be changed for
\begin{equation}
\Delta(\delta m_u^2)\sim \Delta(\delta m_h^{(t)2}),\end{equation}
where $\Delta(\delta m_h^{(t)2})$ indicates the other two gauge interaction induced soft mass for top squarks. Such a change corresponds to the coupling
\begin{equation}\label{small}
\frac{\Delta \lambda_u}{\lambda_u}\sim\frac{\alpha_3^2}{\lambda_u^2}\frac{(F_3/M_3)^2}{(F_2/M_2)^2}\frac{y_t^2}{8\pi^2}\log\frac{M_2}{m_{\widetilde{t}_L}}\ll1
\end{equation}
from the estimation~(\ref{estimation}). The major fine tuning does not depends on top squark mass.

It is known that the Higgs mass in gauge mediation is hard to be $125~\textmd{GeV}$ because of the smallness of squark mixing terms which are generated at higher loops~\cite{Carena:1995bx,Arbey:2011ab,Draper:2011aa}. 
The large splitting in masses~(\ref{mass patten}) and  $\mu$-term~(\ref{mu}) induced mixing term gives us a possibility to generate  a heavy Higgs bosons in gauge mediation. 
Neglecting the all of the trilinear soft supersymmetry breaking terms, the corrections to the lightest Higgs mass is 
\begin{equation}\label{mu correction}
\delta m_h^2=\frac{3\sqrt{2}}{2\pi^2}G_Fm_t^4\left[\log\bigg(\frac{m_{\widetilde{t}_L}m_{\widetilde{t}_R}}{m_t^2}\bigg)+\frac{(\mu\cot\beta)^2}{m_{\widetilde{t}_L}m_{\widetilde{t}_R}}\bigg(1-\frac{1}{12}\frac{(\mu\cot\beta)^2}{m_{\widetilde{t}_L}m_{\widetilde{t}_R}}\bigg)\right],
\end{equation}
where the first and second terms on the right hand side come  from the mass splitting and mixing, respectively. 
The dominant contribution to the Higgs mass is from the logarithmic corrections of the top squark mass, the heavier the top squark the heavier the Higgs boson. 
From eq.~(\ref{estimation}) we have
\begin{equation}
\mu\sim\frac{\lambda_u\lambda_d}{16\pi^2}\frac{F_2}{M_2}\sim\left(\frac{g^2}{16\pi^2}\right)^2\frac{F_2}{M_2}\sim\frac{\alpha_2}{4\pi}m_{\widetilde{e}_L}.
\end{equation}
Neglecting higher order term,  the $\mu$-term induced correction can be estimated by
\begin{equation}
\frac{3\sqrt{2}}{2\pi^2}G_Fm_t^4\left[\bigg(\frac{\alpha_2}{4\pi}\bigg)^2\frac{m_{\widetilde{e}_L}}{m_{\widetilde{e}_R}}\cot^2\beta\right].
\end{equation}
For sufficient heave doublet sparticles, together with the sizable  $\mu$ correction, the Higgs boson can be as heavy as $125~\textmd{GeV}$.

Notice that the above model~(\ref{gauge mediation details}) is only an example, it is weakly coupled and easy to  show the  basic features of the scenario of heavy doublet sparticles. In fact the hidden sector physics can be much more complicate or even strongly coupled.  The ten parameters in the gauge mediation are all free parameters unless the underlying physics is detcovered.

Unlike the left handed sparticles, the right handed squarks and sleptons are not necessary to be heavy to contribute the electroweak symmetry breaking. They can be light to be able to produce  and observe at the LHC in future. The predictions and implications of the model are discussed in the followings.

Since left handed sparticles are all heavy in the scenario, the physics below the \textmd{TeV} scale is simpler than those in the MSSM. There are only four relevant parameters
\begin{equation}
m_{\widetilde{g}},m_{\widetilde{b}},M_3\textrm{ and } M_1.
\end{equation}
Where $m_{\widetilde{g}}$ and $m_{\widetilde{b}}$ are gluino and bino masses, respectively. 
$M_3$ and $M_1$ are gauge mediated masses for $SU(3)$ and $U(1)$ gauge groups, they are defined by
\begin{equation}
M_{1,3}^2=\left(\frac{\alpha_{1,3}}{4\pi}\right)^2A_{1,3}.
\end{equation}  
Although the  $SU(2)$ gaugino mass $m_{\widetilde{W}}$ does not contribute to the Higgs mass square matrix~(\ref{higgs mass matrix}), it is also believed to be as large as the doublet sparticle masses as  in the  example~(\ref{gauge mediation details}). For the Higgs mixing parameter $\mu$, it can be above or below the \textmd{TeV} scale in different cases. For simplicity it is assumed that the Higgsinos are heavy. Due to the largeness of the left handed sparticle masses, the mixings between left and right handed sfermions can be neglected.

The light sfermions are superpartners of  right handed fermions $\widetilde{u_i}_R$, $\widetilde{d_i}_R$ and $\widetilde{l_i}_R$. Their masses are given by
\begin{eqnarray}
m_{\widetilde{u}_R}^2 &=& M_3^2+(\frac{2}{3})^2M_1^2, \\
  m_{\widetilde{d}_R}^2 &=& M_3^2+(\frac{1}{3})^2M_1^2, \\
  m_{\widetilde{l}_R}^2 &=& M_1^2.
\end{eqnarray}
There is a simple relation among these masses
\begin{equation}\label{mass relation}
m_{\widetilde{u}_R}^2=m_{\widetilde{d}_R}^2+\frac{1}{3}m_{\widetilde{L}_R}^2
\end{equation}
that can be tested as an evidence of gauge mediation if all of them are discovered.

Furthermore to distinguish the heavy doublet scenario with the ordinary gauge mediation in which colored superpartners are generally considered heavier then those without colors, we have to measure the decay widths of the sfermions. In the usual supersymmetric models, superpartners are generally mixed together if they have the same Lorentz and gauge quantum numbers. The  properties of physical particles depend on many unknown parameters that we loss the accurate prediction to them.  For example  in the MSSM both  gauge and Yukawa interactions contribute to the slepton  decay $\widetilde{l}\rightarrow l\chi_0$, where $\chi_0$ is the lightest supersymmetric particle (LSP),
the decay width depends on many unknowns.  On the other hand, if the doublet sparticles, $SU(2)$ gauginos and Higgsinos  are all taken to be heavy,  the only decay channel is
\begin{equation}
\widetilde{l}_R\rightarrow l_R+\widetilde{b}
\end{equation}
though supersymmetric $U(1)$ gauge interaction. The decay width is determined by the mass of slepton and bino, neglecting the lepton masses the decay width is given by
\begin{equation}\label{Glr}
\Gamma(\widetilde{l}_R\rightarrow l\widetilde{b})=\frac{g'^{2}m_{\widetilde{l}_R}}{8\pi}\big(1-\frac{m_{\widetilde{b}}^2}{m_{\widetilde{l}_R}^2}\big)^2.
\end{equation}
Similarly the decay width of the right handed down type squarks for $m_{\widetilde{g}}>m_{\widetilde{d}_R}$ in the small quark mass limit is
\begin{equation}\label{Gdr}
\Gamma(\widetilde{d}_R\rightarrow d\widetilde{b})=\frac{1}{3}\frac{g'^{2}m_{\widetilde{d}_R}}{8\pi}\big(1-\frac{m_{\widetilde{b}}^2}{m_{\widetilde{d}_R}^2}\big)^2.
\end{equation}

Precise measurement of the masses and widths of the $\widetilde{l}_R$ and $\widetilde{d}_R$  can justify if the scenario is correct or not using the derived identity from eqs.~(\ref{Glr},\ref{Gdr})
\begin{equation}\label{prediction}
m_{\widetilde{l}_R}^2\big[1-\sqrt{\frac{8\pi\Gamma(\widetilde{l}_R)}{g'^2m_{\widetilde{l}_R}}}\big]=m_{\widetilde{d}_R}^2\big[1-\sqrt{3\frac{8\pi\Gamma(\widetilde{d}_R)}{g'^2m_{\widetilde{d}_R}}}\big]=m_{\widetilde{b}}^2.
\end{equation}
It can also determine the mass of the LSP if the first equality is confirmed by experiments.

\section{Discussions and Conclusions}
The  particles in the $SU(2)$ doublets are special in the Standard Model and its extensions. They violate the parity and participate the $SU(2)$ gauge interactions. In addition  the Higgs fields which break the electroweak gauge symmetry are also in the $SU(2)$ doublet. It is very possible that there are internal connections between the properties of the doublets and the electroweak symmetry breaking. An example of such a connection is established in this article.

In gauge mediation masses of superparticles are determined by their gauge quantum numbers.  Assuming that the supersymmetry breaking in the $SU(2)$ gauge group is much larger than the other gauge interactions, many problems can be solved. There are light squarks especially the right handed top squarks that may be discovered by the LHC.   In the usual mass patten in gauge mediation in which colored superparticles are considered to be heavier than those without colors, the LHC observable squarks can not contribute enough quantum corrections to lift the Higgs mass. In additional to the light $\widetilde{t}_R$ we also have heavy enough $\widetilde{t}_L$ to contribute the corrections.

Even though the electroweak symmetry breaking is achieved by a considerable fine tuning, but it is less dependent on the top squark mass as analyzed in the previous section.   We can always choose the four parameters in the Higgs sector to obtain a small negative eigenvalue of   the Higgs mass matrix~(\ref{higgs mass matrix}). The mass of the possibly observed top squark in future only affect the chosen parameters small relative quantities~(\ref{small}).

The degeneracy of superpartners with the same quantum numbers   and the relation~(\ref{mass relation}) are unambiguous  test for gauge mediation. We can use eq.~(\ref{prediction}) to justify if the heavy doublet scenario is correct or not when the $\widetilde{d}_R$ and $\widetilde{l}_R$ are discoveried in the 14~\textmd{TeV} run of the LHC.

I would like to thank  Chun.~Liu, Minkai.~Du, Jiashu.~Lu, Shuo.~Yang and Zhenhua.~Zhao for very fruitful  discussions. This work
was supported in part by the National Natural Science Foundation of
China under nos. 11075193 and 10821504, and by the National Basic
Research Program of China under Grant No. 2010CB833000.

\end{document}